# ENGAGING LEARNING ANALYTICS IN MOOCS: THE GOOD, THE BAD, AND THE UGLY

**Mohammad Khalil, Behnam Taraghi & Martin Ebner**
*Educational Technology, Graz University of Technology (Austria)*

## Abstract

Learning Analytics is an emerging field in the vast areas of Educational Technology and Technology Enhanced Learning (TEL). It provides tools and techniques that offer researchers the ability to analyze, study, and benchmark institutions, learners and teachers as well as online learning environments such as MOOCs. Massive Open Online Courses (MOOCs) are considered to be a very active and an innovative form of bringing educational content to a broad community. Due to the reasons of being free and accessible to the public, MOOCs attracted a large number of heterogeneous learners who differ in education level, gender, and age. However, there are pressing demands to adjust the quality of the hosted courses, as well as controlling the high dropout ratio and the lack of interaction. With the help of Learning Analytics, it is possible to contain such issues. In this publication, we discuss the principles of engaging Learning Analytics in MOOCs learning environments and review its potential and capabilities (the good), constraints (the bad), and fallacy analytics (the ugly) based on our experience in last year's.

## 1. Introduction

Since 2008, Massive Open Online Courses (MOOCs) have shown significance and potentiality to scale education in distance learning environments. The benefits shine when thousands of students can participate in a course that a normal classroom cannot hold. Due to many reasons of being free, available to the public and require no predefined level of participation, MOOCs attracted a large number of learners from all over the world regardless their educational background, gender or age. Institutions of Higher Education (HE) start to think seriously of adopting MOOCs and make use of Open Educational Resources (OER) principles. Comparatively, famous MOOC-platform such as Coursera was established by Stanford University, and edX by the Massachusetts Institute of Technology and Harvard. Both platforms provide various courses to university students. Besides, MOOCs are not only preserved to university and college participants, but also for primary school children, such as courses provided by the Austrian MOOC provider, iMooX (www.imoox.at).

Typically, MOOCs are based on video lectures, multiple-choice quizzes or peer-review assessments, discussion forums and documents (Khalil & Ebner, 2016c; Lackner, Ebner & Khalil, 2015). Lessons are delivered on a weekly basis, and students commit to attend during the week. Additionally, students can solve assignments and then share and discuss their views in forums or social media networks. Further, teachers post questions and can communicate with students toward creating a domain of presence (Khalil & Ebner, 2013). Nevertheless, frequent studies and reports complain about the low completion rate, lack of interaction (Lackner, Ebner & Khalil, 2015), keeping the learners motivated, engagement issues, and last but not least cheating and gaming the MOOC systems (Khalil & Ebner, 2015a; Khalil, Kastl & Ebner, 2016). As a result, mining student actions on distance learning environments makes the job easier for educationists and researchers to maintain learner behaviors and explain such concerns.

An inclusion and exploration of the term "Big Data" in the education field emerged recently. Two main research communities oriented with respect to discovering new meaning of educational datasets activities: the Educational Data Mining and the Learning Analytics communities (Papamitsiou & Economides, 2014). In this paper, the focus will be mainly on Learning Analytics. We will discuss the potential and capabilities as well as the constraints and the negative sides of the field with strong focus on MOOCs. These criteria are established based on our experience in the last couple of years of implementing Learning Analytics prototypes and strategies in the Austrian iMooX platform.

## 2. Learning analytics potentiality in MOOCs (the good)

Analyzing student data on online environments in order to reveal hidden patterns and discover paradigms of activities is called Learning Analytics. In 2011, the Society for Learning Analytics and Research defines it as "… the measurement, collection, analysis and reporting of data about learners and their contexts, for purposes of understanding and optimizing learning and the environment in which it occurs". The needs for Learning Analytics emerged to optimize learning and benchmark the learning environments. Khalil and Ebner (2015b, 2016c) discussed the various promises of employing Learning Analytics in MOOCs platforms. Another recent study by Khalil and Ebner (2016b) about surveying Learning Analytics techniques from 2013 to 2015 shows that the combination of Learning Analytics and MOOCs related-topic scored the highest number of citations in Google Scholar (http://scholar.google.com) during that period.

Online distance learning environments such as MOOCs provide a rich source of knowledge mining opportunity. By logging mouse clicks, forums activity, quiz performance, login frequency, time spent on tasks and tracking videos interactivity, Learning Analytics researchers can build an enormous amount of data logs. This database of information, if interpreted appropriately, can help researchers from diverse disciplines of computer science, pedagogy, statistics, and machine learning…etc., to intervene directly toward student success. Benefits of Learning Analytics in MOOCs are limitless. In the following, we list the primary benefits of applying Learning Analytics in MOOCs:

- *Prediction*: One of the most popular objectives performed by both Learning Analytics and Educational Data Mining. Techniques are used to predict when a participant is expected to drop from an online course. This could be done by analyzing a student behavior, exam performance, and video skips. Storing numerous records of previous students' activities based on specific modules help researchers predict the prospective action, such as dropping out of a course or detecting students at-risk. Additionally, Learning Analytics is used in predicting performance and motivation (Edtstadler, Ebner & Ebner, 2015). Further forecasting about video watching on a course and relative activity in discussion forums is feasible to be investigated.

- *Recommendation*: Actions on MOOC platforms can be mined for recommendation purposes. An example is when a MOOC provider recommends learning materials to students based on their previous registered courses. In addition, recommendations can be generated to suggest a student answering a specific question in discussion forums.

- *Visualization*: Through Learning Analytics, tracking previously mentioned actions creates a lot of records. Visualizations can be presented to participants via dashboards. Verbert and her colleagues (2014) discussed that dashboards support awareness, reflection and sense-making. On the other hand, analyzing data through visualizing them into plots supports researchers to reveal patterns (Khalil & Ebner, 2016c) and provides feedback and reflection to MOOC participants at the end.

- *Entertainment*: Gaming tools were considered as a Learning Analytics technique in Khalil and Ebner (2016b) work. The survey illustrates how gamification makes learning in MOOCs more entertaining which results in an increased motivation and completion rate among students. Such tools can be badges (Wüster & Ebner, 2016), reward points, progress bars or colorful gauges.

- *Benchmarking*: Benchmarking is a learning process which evaluating courses, videos, assignments, and MOOC platforms are attainable using Learning Analytics. Hence, we can identify learning difficulties as well as weak points in the online courses or stalling segments in video lectures. Accordingly, constructive feedback is generated which concludes into an enhanced educational system.

- *Personalization*: Learners can shape their personal experience in a MOOC. Developers through different types of Learning Analytics techniques (Khalil & Ebner, 2016b) can build a set of personalized items in the MOOC platform. For example, a student can favorite a part of a video or bookmark an article or a document. Further, (s)he can customize notifications and add annotations in videos.

- *Enhance Engagement*: Engagement has recently been an attracting topic in MOOCs. Employing Learning Analytics through data mining techniques such as clustering was used in (Kizilcec, Piech & Schneider, 2013; Khalil, Kastl & Ebner, 2016). Expected results are grouping participants into a subpopulation of students or classifying interactions in videos, assignments and quizzes for reasons of future interventions in MOOC designs or studying the needs of the students' catalogue.

- *Communication Information*: Learning Analytics involves collecting data from sources and processes them. It is further used to report information in a form of statistical analysis to different MOOC stakeholders. Similar to web analytics, students can check their activities and review general statistics using dashboards, for example. In addition, teachers and decision makers can build an overview about MOOC using descriptive statistics.

- *Cost Saving*: Since Learning Analytics provides tools of data analysis, it opens the doors for a broad examination services which makes it possible to determine weak sections of a MOOC. Therefore, decision makers can allocate resources effectively.

## 3. The negative side of learning analytics in MOOCs (the bad)

Despite the fact that Learning Analytics achieves several benefits when it is applied to education data stream, rising constraints have been identified lately (Papamitsiou & Economides, 2014; Khalil & Ebner, 2015b). The large-scale of data collection and processes drives Learning Analytics to questions related to privacy and ethical issues. An atmosphere of uncertainty among practitioners of Learning Analytics as well as decision makers decelerates its steep growth (Drachsler & Greller, 2016). Through our experience, we encourage educational organizations to adopt the security model CIA, which stands for Confidentiality, Integrity, and Availability. In this section, we list major concerns of implementing Learning Analytics in MOOC platforms:

- *Security:* The stored records of students in databases that belong to Learning Analytics applications represent the heart of their private information. Thus, maintaining database configuration is not always considered by organizations. As a result, breaches of confidential information are possible to happen.
- *Privacy:* Learning Analytics can reveal personal information of learners. MOOC datasets may hold sensitive information such as emails, names or addresses. Privacy has been considered as a threat in Learning Analytics (Papamitsiou & Economides, 2014) and as a constraint (Khalil & Ebner, 2015b). Different solutions can be proposed such as anonymization approach (Khalil & Ebner, 2016a), encryption, or increasing restrictions.
- *Ownership:* Questions related to "who owns the analyzed data of MOOCs" can emerge anytime. Participants like to keep their information confidential, but at the same time, consent policy is essential to ensure transparency. Further, MOOC providers are encouraged to delete or de-identify personal information of their participants.
- *Consent:* Related to ownership of data. Not every MOOC provider clearly declares the usage of students' data. Policies with legislation frameworks should include rules of a collection of personal information and a description of information usage, such as research purposes or third party information selling.
- *Transparency:* Secret processes can hide unfair decision making when analytics is applied on educational datasets (Sclater, 2014). By the same token, when Learning Analytics is applied on MOOCs, providers need to disclose their approach to collecting, analyzing and using of participants' data. At the same time, a point of balance should be made when the Learning Analytics algorithms or tools are proprietary. Sclater argued different code of practices regarding transparency.
- *Storage:* As long as MOOCs are open to the public, a single course can attract thousands of students. Storing big data could be costly, overloaded, and complex as well as hard to manage. Furthermore, according to the European Directive 95/46/EC[1], personal data needs to be stored no longer than necessary.

## 4. The dark side of learning analytics in MOOCs (the ugly)

Looking for the quality of data is an important factor in Learning Analytics. However, when data records have incomplete segments or polluted information, then Learning Analytics is negatively affected. Moreover, getting a holistic overview of students in online courses cannot only be harvested through their left traces on MOOCs. Are there any guarantees of the Learning Analytics results? What about the accuracy? In this section, we summarize some of the worst-case results that Learning Analytics can produce by employing it in MOOCs.

- *False Positives:* Making decisions, either by analysts or directors, based on a small subset of data could lead to fast judgments and hence trigger "false positives". Consequently, the accuracy of any forthcoming decision in a MOOC system will be influenced. For instance, if a group of students were "gaming the system" and an analyst builds a prediction model for all students based on MOOC indicators fulfillment, then a false positive action is triggered. As a matter of fact, Learning Analytics is not only based on numbers and statistics. Judgments and opinions of researchers play a major role. We always see flounce on MOOCs discussion forums activity and its correlation with performance. Some researchers approved that more social activity in forums is reflected positively on performance while others go against this theory. In the light of that, Learning Analytics is not always accurate.
- *Fallacy Analytics:* Analytics could fail and thus, mistaken interventions or predictions occur. Failures could happen during the main processes of Learning Analytics cycle. Wrong actions in collecting data from MOOCs, errors in processing or filtering and mistaken interpretation of data are possible scenarios of fallacy analytics. Additionally, presenting the results through visualizations might also be within the same page. Visualizations are a great way to report information, but playing with scales or using 3D figures might be tricky to the end user (student, teacher, decision maker). Fallacy analytics may be accidental and not intentional; however, using interpreted data based on fallacy analytics can be dangerous to different stakeholders and uneconomical to the MOOC business. Fallacy analytics through misuse of statistics as a method of Learning Analytics corrupts and pollutes research records as well as wastes the time and energy of other researchers (Gardenier & Resnik, 2002).

---

[1] http://eur-lex.europa.eu/LexUriServ/LexUriServ.do?uri=CELEX:31995L0046:en:HTML (last visited: March, 2016)

- *Bias:* Learning Analytics can show significant results of prediction and recommendation. It can also prove hypotheses such as the relation between activity in discussion forums and performance or watching videos and passing MOOCs. Collected data "could feel" that, but this actually returns to the intention desire of the researcher or decision maker. The bias towards a certain hypothesis and the inner determination of proving a theory of students' data leads to biased Learning Analytics.

- *Meaningful data:* Papamitsiou and Economides (2014) mentioned that Learning Analytics mostly uses quantitative research results. Qualitative methods have not yet shown significant results. Learning Analytics can be ineffective and waste of efforts if meaningful data is hard to extract. Dringus (2012) argued two main points regarding meaningful data in Learning Analytics: 1) if the data collected has no impact on improving or changing education. 2) if the data has no meaningful evidence such as lack of clarity about what to measure to get meaningful information.

## 5. Conclusion

Learning Analytics provides various tools and to optimize learning. In this paper, we reviewed the principles of engaging Learning Analytics in Massive Open Online Courses (MOOCs). We discussed the capabilities (the good), the dilemmas (the bad) and the out of the bound situations (the ugly).

*Figure 1. The advantages and disadvantages of Learning Analytics in MOOCs*

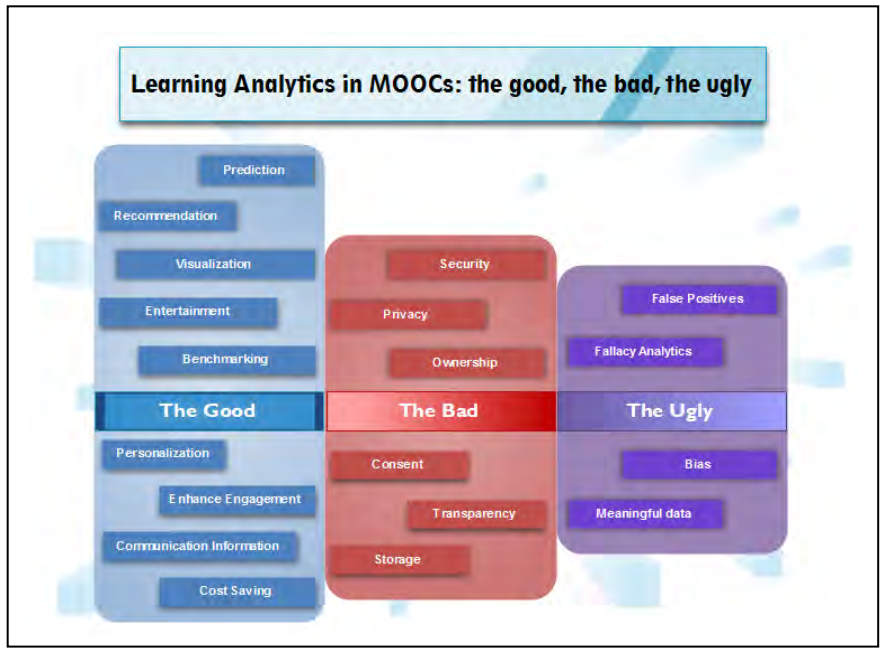

Figure 1 summarizes our results. Generally speaking, MOOCs and Learning Analytics imply high potentiality. Nevertheless, a code of practice should be considered by all stakeholders in order to carry out the optimum outcomes.